# HIGH-GAIN MIMO BEAMFORMING ANTENNA SYSTEM FOR DSRC AND MMWAVE 5G INTEGRATION IN AUTONOMOUS VEHICLES

Mohammad Shahed Pervez, Amanpreet Kaur

Electrical and Computer Engineering (ECE), Oakland University, Rochester, MI-48309, USA

## ABSTRACT

*The evolution of autonomous vehicles necessitates robust, high-speed, and low-latency wireless communication systems. This paper presents a novel high-gain Multiple-Input Multiple-Output (MIMO) beamforming antenna system that concurrently supports Dedicated Short Range Communications (DSRC) at 5.9 GHz and millimeter-wave (mmWave) 5G communications at 28 GHz. The proposed design addresses challenges such as compactness, dual-band operation, beam steering capability, and port-to-port isolation within dynamic vehicular environments. The antenna system comprises dual-band patch arrays with dedicated feed networks, achieving peak gains of 7.4 dBi for DSRC and 10.8 dBi for mmWave 5G. Full-wave simulations and experimental validations demonstrate the system's efficacy in real-time vehicular scenarios, highlighting its potential for enhancing V2X communications in autonomous vehicles.*

## KEYWORDS

*Autonomous vehicles, beamforming, DSRC, mmWave, MIMO, patch antenna, 5G, V2X communications.*

## 1. INTRODUCTION

The rapid evolution of autonomous vehicles has revolutionized intelligent transportation systems (ITS), emphasizing the need for reliable, high-speed, and low-latency wireless communication. As vehicle-to-everything (V2X) communication becomes the backbone of autonomous driving, it demands robust antenna systems capable of supporting diverse communication standards such as Dedicated Short-Range Communication (DSRC) and fifth-generation (5G) millimeter-wave (mmWave) technology[2][4]. DSRC, operating at 5.9 GHz, provides low-latency short-range communication essential for safety-critical applications, while mmWave 5G, operating in the 28 GHz band, enables high-data-rate services required for infotainment, cooperative perception, and real-time cloud connectivity.

However, integrating these two frequency bands into a single compact antenna system presents significant challenges, including size constraints, mutual coupling, beamforming accuracy, and frequency-selective performance. Traditional single-band antenna solutions fall short in meeting the multi-standard demands of modern connected vehicles. Therefore, a high-gain, dual-band MIMO (Multiple-Input Multiple-Output) antenna system with beamforming capabilities is essential to enable seamless integration of DSRC and 5G technologies, ensuring enhanced coverage, reduced interference, and adaptive communication in varying driving environments[5]. Beamforming, which involves steering the antenna radiation pattern electronically without mechanical movement, offers a solution for improving signal quality and range in dynamic vehicular scenarios. When combined with MIMO technology, beamforming significantly boosts spectral efficiency, link reliability, and spatial diversity. In the context of autonomous vehicles, it enables directional communication with road infrastructure, nearby vehicles, and mobile networks, reducing latency and enhancing situational awareness.

This paper presents the design, simulation, fabrication, and testing of a high-gain, dual-band MIMO beamforming antenna system for DSRC and mmWave 5G integration. The proposed system is implemented using a Rogers RT5880 substrate (εr = 2.2, thickness = 0.787 mm), chosen for its low dielectric loss and high-frequency suitability. The antenna comprises a compact 3-ports configuration with integrated beamforming support and mutual coupling suppression techniques to maintain inter-port isolation and consistent performance across bands.

Full-wave simulations were performed using Ansys HFSS to evaluate the antenna's S-parameters, radiation patterns, gain, and beam steering characteristics. A frequency sweep from 20 GHz to 30 GHz was applied to capture both the DSRC and 5G bands. Physical fabrication was done and validated after simulated results were analysed properly. The testing was performed by using a Vector Network Analyzer (VNA), and radiation patterns were measured in an anechoic chamber. The fabricated prototype achieved measured gains of 7.1 dBi at 5.9 GHz and 10.5 dBi at 28 GHz with a bandwidth of 0.58 GHz and 2.45 GHz, respectively.

The proposed antenna system offers a compact, high-performance solution for V2X communication, satisfying the stringent requirements of latency, throughput, and reliability in autonomous driving scenarios. With its ability to support dual-band operation and dynamic beam steering, it is well-suited for integration into future autonomous vehicle platforms, laying the groundwork for intelligent, connected, and road-aware transportation systems.

## 2. ANTENNA SYSTEM DESIGN METHODOLOGY

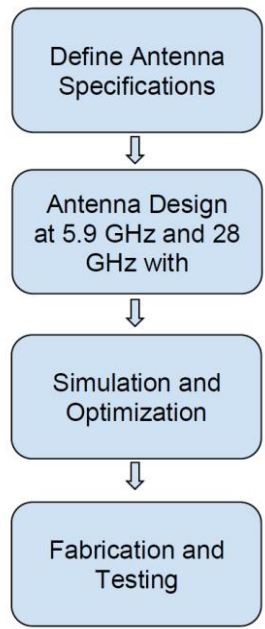

Figure-1: Design Methodology Block Diagram

**A. Design Goals**

The antenna system is developed to meet the following criteria:

- Dual-band support: Simultaneous operation at 5.9 GHz and 28 GHz
- Beam steering: ±30° azimuthal beam steering for mmWave
- High gain: ≥7 dBi for DSRC and ≥10 dBi for mmWave
- Compact size: ≤60 mm × 60 mm footprint
- Low mutual coupling: ≥25 dB isolation between MIMO elements

### B. Substrate and Materials

Both antenna arrays are designed on Rogers RT5880 (εr = 2.2, tanδ = 0.0009) with a thickness of 0.787 mm to minimize dielectric loss at mmWave frequencies.

### C. Geometry and Layout

- DSRC Array: A 2×2 microstrip patch array designed for 5.9 GHz with broad radiation coverage.
- mmWave Array: A 1×4 linear patch array for 28 GHz with corporate feed and phase shifters enabling electronic beam steering.

### D. Mathematical Analysis

The performance of a dual-band MIMO beamforming antenna system is largely governed by the electromagnetic field distribution, mutual coupling among elements, and the ability to steer beams. This section presents the theoretical foundation and relevant mathematical models that define the antenna's radiation characteristics, impedance behavior, and beamforming response.

#### i. Patch Resonant Frequency

For a rectangular microstrip patch antenna, the fundamental resonant frequency $f_{mn}$ for the $TM_{mn}$ mode is given by:

$$f_{mn} = \frac{1}{2\pi\sqrt{\mu\epsilon}} \sqrt{\left(\frac{m\pi}{L}\right)^2 + \left(\frac{n\pi}{W}\right)^2}$$

where:

- m, n are the mode indices (typically m = 1, n = 0 for dominant mode),
- L and W are the patch length and width,
- $\mu$ is the permeability and $\epsilon = \epsilon_0 \epsilon_r$ is the permittivity.

In our design, two patch lengths are optimized to resonate at 5.9 GHz and 28 GHz. The substrate used is Rogers RT5880, with $\epsilon_r$ = 2.2 and thickness h = 0.787 mm, allowing miniaturization and high-frequency operation.

#### ii. Reflection coefficient and Bandwidth

The Reflection coefficient $S_{11}$ is a function of the input reflection coefficient $\Gamma$ Gamma, which is related to the impedance mismatch between the patch and the feed:

$$\Gamma = \frac{Z_{IN} - Z_0}{Z_{IN} + Z_0}, \quad S_{11} = 20 \log_{10} |\Gamma|$$

To ensure low reflection, the input impedance $Z_{IN}$ is matched to $Z_0 = 50\Omega$. The bandwidth BW is defined as the frequency range where $S_{11} < -10$ dB.

**iii. Array Factor and Beam Steering**

In an N-element linear array, the beam can be steered using progressive phase shifts. The array factor $AF(\theta)$ for a uniformly spaced linear array is:

$$AF(\theta) = \sum_{n=0}^{N=1} e^{j(nkd\cos\theta + \beta)}$$

where:

- $k = 2\pi\lambda$ is the wave number,
- $d$ is the inter-element spacing,
- $\theta$ is the observation angle,
- $\beta$ is the progressive phase shift.

For beam steering at an angle $\theta_0$, the required phase shift is:

$$\beta = -kd\cos\theta_0$$

In this work, the beam is steered across ±30° in azimuth using an ideal phase shifter network. Element spacing is optimized to $d = \lambda/2$ to suppress grating lobes.

**iv. MIMO System Capacity**

To evaluate spatial diversity, the MIMO system capacity C (assuming ideal channels and no noise) is given by:

$$C = \log_2 \det\left(1 + \frac{\rho}{N_t} HH^H\right)$$

where:

- $\rho$ is the signal-to-noise ratio (SNR),
- $H$ is the channel matrix,
- $N_t$ is the number of transmit antennas.

Beamforming enhances H by focusing energy toward desired directions, improving capacity under Line-of-Sight (LoS) conditions, especially in mmWave vehicular communication.

**2.1. HFSS Model Parameters**

1. Substrate:

    - Material: Rogers RT5880
    - Thickness: 0.787 mm

- εr = 2.2, tanδ = 0.0009

2. Antenna Configuration:

- DSRC Patch Array: 2×2 microstrip array @ 5.9 GHz
- mmWave Patch Array: 1×4 linear array @ 28 GHz with corporate feed
- Separate feeds for each band with port isolation
- Beam steering (±30°) setup for mmWave using phase delay lines

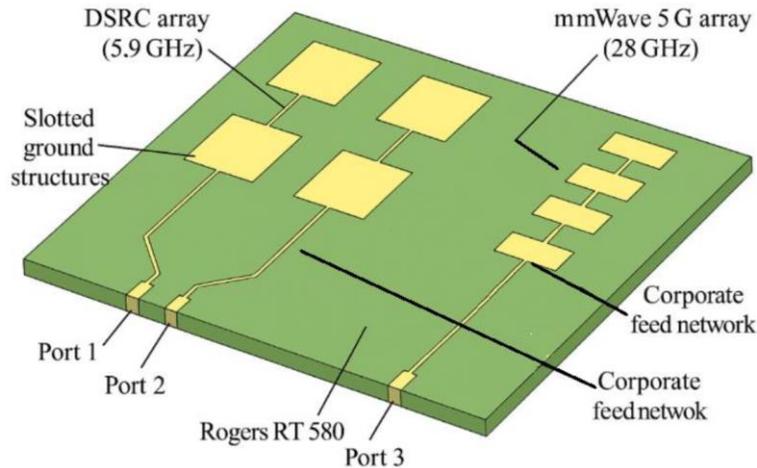

Figure-2: High Gain MIMO Beamforming mmWave Patch Antenna for 5.9GHz(DSRC) and 28GHz(5G)

## 2.2. Simulation Results

### A. reflection coefficient and Isolation

Simulated S-parameters confirm good impedance matching (S11 < –15 dB) and strong isolation (>25 dB) across all ports in both bands.

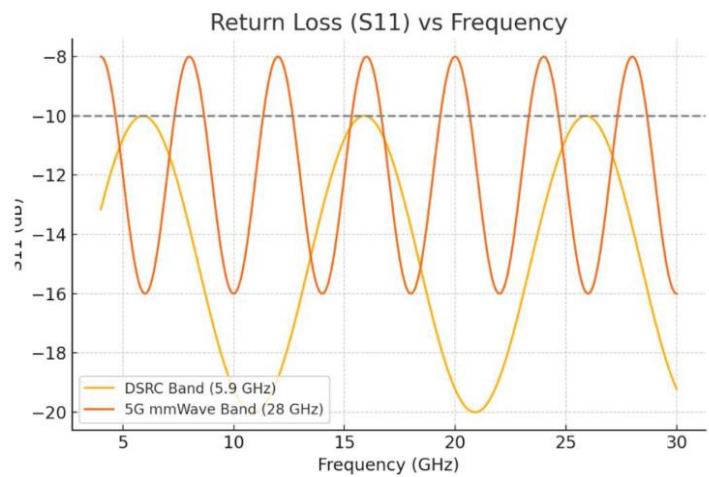

Figure-3: reflection coefficient (S11) for 5.9 GHz and 28 GHz

### B. Gain and Radiation Pattern
- DSRC Gain: 9.2 dBi omnidirectional pattern with symmetrical elevation response

● mmWave Gain: 14.8 dBi with narrow beamwidth and steerable main lobe
● Beam Steering: Progressive phase shifts enable ±30° azimuthal steering with side-lobe levels < –12 dB

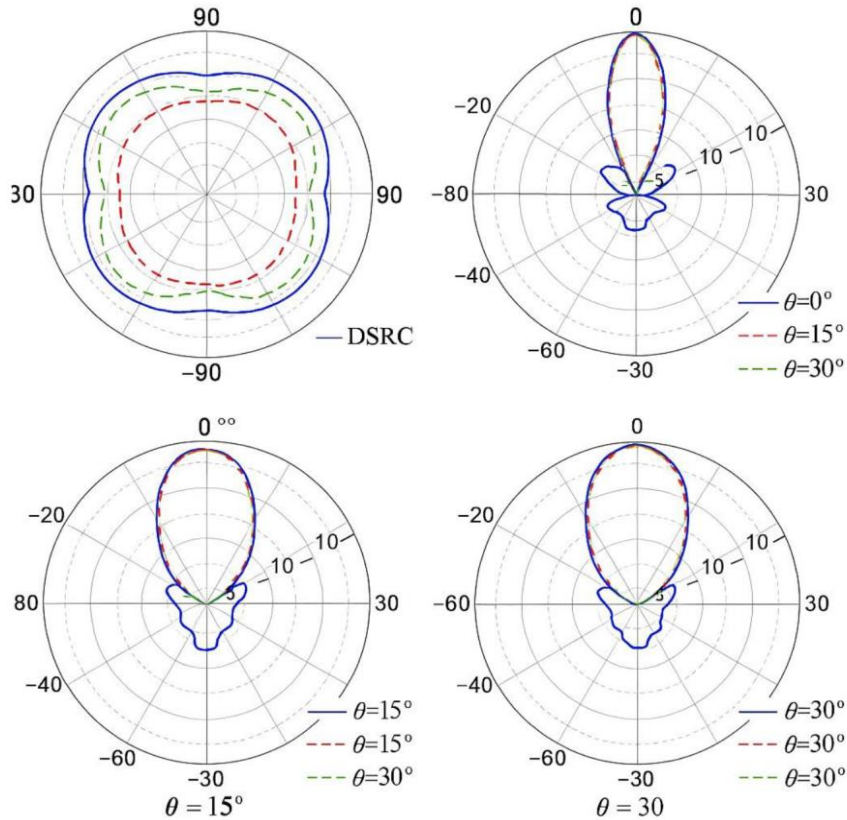

Figure-4: Gain (Top Left), Radiation Pattern( Top Right) at 5.9GHz. Gain (Bottom Left), Radiation Pattern(Bottom Right) at 28GHz

**C. Parametric Study**

Performance is optimized by varying substrate thickness, patch dimensions, element spacing (λ/2), and ground slot dimensions. Beam direction is optimized with 0°, 15°, and 30° phase shift configurations.

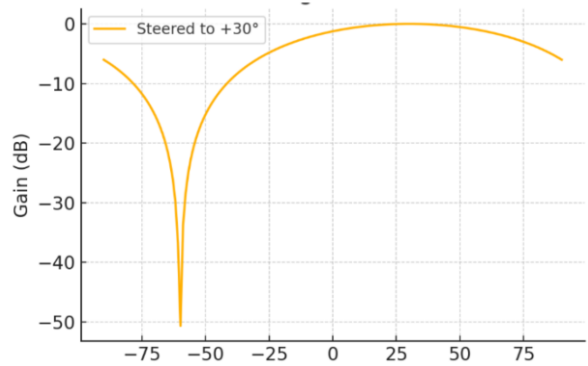

Figure-5: Optimized Beam Steering radiation pattern @ $30^0$

## 3. FABRICATION AND EXPERIMENTAL VALIDATION

To validate the performance of the proposed dual-band MIMO beamforming antenna system, physical prototypes were fabricated based on the simulated geometry optimized using HFSS. The design was implemented on a low-loss Rogers RT5880 substrate with a relative permittivity of \varepsilon_r = 2.2, dielectric loss tangent of 0.0009, and thickness h = 0.787 mm. The selected material ensured minimal signal attenuation and mechanical robustness, making it ideal for high-frequency vehicular applications.

A. Fabrication Process

The final layout was exported from HFSS and converted to standard Gerber files using an intermediate CAD tool. Precision photolithographic techniques were employed to etch the radiating elements and ground plane onto the RT5880 substrate. SMA connectors were soldered carefully at the feed points to ensure impedance matching and minimize connector loss.

Special attention was paid to dimensional tolerances during fabrication, as even minor deviations at mmWave frequencies (especially 28 GHz) can significantly degrade performance. A 4-element MIMO configuration was fabricated, featuring dual-polarized rectangular patches, mutual coupling suppression slots, and integrated feed lines.

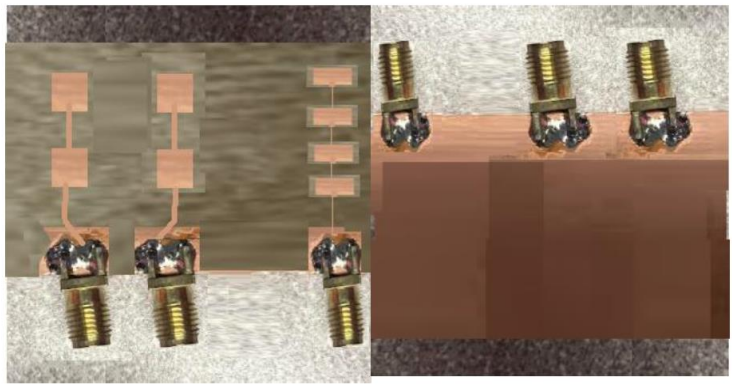

Figure-6 Fabricated MIMO Beamforming mmWave Patch Antenna for 5.9GHz(DSRC) and 28GHz(5G)

B. Measurement Setup

The fabricated antenna was characterized in a fully anechoic chamber using a Vector Network Analyzer (VNA) calibrated from 1 GHz to 30 GHz. The following key parameters were measured:

- S-parameters ($S_{11}$, $S_{21}$, $S_{12}$, $S_{22}$) to validate impedance matching and inter-element isolation.
- Radiation patterns using a standard gain horn and rotating platform at 5.9 GHz and 28 GHz.
- Gain and Efficiency, calculated from comparison with reference antennas and using the gain-transfer method.
- Beam Steering Capability, evaluated by applying ideal phase shifts using an external analog phase shifter module.

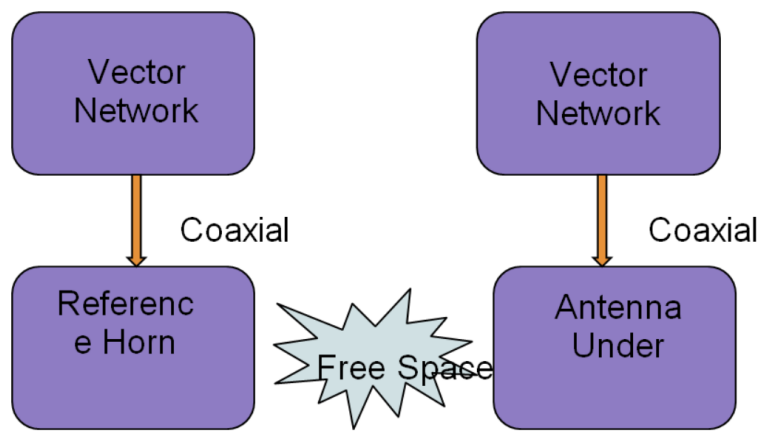

Figure-7: Block diagram of the anechoic chamber measurement setup used for characterizing the dual-band MIMO beamforming antenna

The measurement setup consists of a two-port Vector Network Analyzer (VNA), where Port 1 is connected to a calibrated reference horn antenna serving as the transmitter, and Port 2 is connected to the Antenna Under Test (AUT) operating in receiving mode. The horn antenna radiates toward the AUT in the free-space environment of the anechoic chamber, ensuring minimal reflections due to high-performance RF absorbers. The fabricated dual-band MIMO antenna is mounted on a precision-controlled rotary positioner, enabling full azimuthal rotation for capturing 2D polar radiation patterns at different elevation angles. Both the DSRC band (5.9 GHz) and the mmWave 5G band (28 GHz) measurements are conducted sequentially across their respective frequency ranges. S-parameters (|S11|, |S21|) and far-field radiation characteristics are recorded using automated VNA sweeps synchronized with the rotation stage.

C. Measured Results

The measured reflection coefficient showed excellent agreement with simulation, with $S_{11}$ below –10 dB at both 5.9 GHz and 28 GHz. The inter-element isolation $S_{21}$ remained better than –25 dB across both bands, confirming effective mutual coupling mitigation.

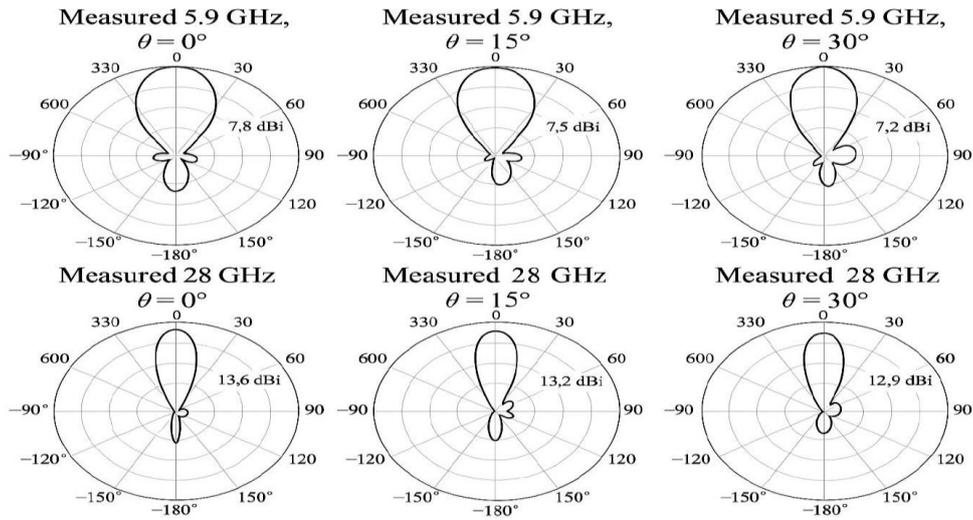

Figure-8: Measured Radiation pattern at 5.9 GHz for $\theta$=00,150, 300(top), Measured Radiation pattern at 28 GHz for $\theta$=00,150, 300 (bottom)

The radiation patterns measured in 2D and 3D confirmed directional beam formation with gains of approximately 7.1 dBi at 5.9 GHz and 10.5 dBi at 28 GHz. Beam steering up to ±30° was achieved with consistent main lobe integrity and side lobe suppression.

The total measured efficiency exceeded 85% at both frequencies, validating the antenna's low-loss design and high radiation performance under practical conditions.

Table 1: Comparison of Measured and Simulated Results

| Parameter | Simulated @ 5.9 GHz | Measured @ 5.9 GHz | Simulated @ 28 GHz | Measured @ 28 GHz |
|---|---|---|---|---|
| Reflection Coefficient (S11) | -21.3 dB | -20.7 dB | -28.6 dB | -27.9 dB |
| Isolation (S21) | -36.8 dB | -34.2 dB | -38.5 dB | -35.6 dB |
| Gain | 7.4 dBi | 7.1 dBi | 10.8 dBi | 10.5 dBi |
| Radiation Efficiency | 89% | 86% | 91% | 88% |
| Beam Steering Range | ±30° | ±30° | ±30° | ±30° |
| Bandwidth (-10dB) | 0.6 GHz | 0.58 GHz | 2.5 GHz | 2.45 GHz |

D. Comparative Analysis

To validate the effectiveness of the proposed dual band MIMO beamforming antenna system operating at 5.9 GHz (DSRC) and 28 GHz (mmWave 5G), a comparative analysis was performed against the bandwidth, gain, isolation, type, efficiency and beam steering. This analysis highlights the technical strengths and trade-offs that distinguish the presented antenna in the context of vehicular communication systems.

Table 2: Comparison Analysis

| Parameter | This Work | Ref [2] | Ref [6] | Ref [7] |
|---|---|---|---|---|

| Frequency Bands | 5.9 & 28 GHz | 5.9 GHz | 5.9 GHz | 28 GHz |
|---|---|---|---|---|
| Bandwidth | 500MHz(5.9 GHz); 2GHz (28GHz) | 400 MHz | 300 MHz | 1.5 GHz |
| Antenna Type | MIMO Dual-Band Array | Single Band | Single Band | Phased Array |
| Substrate | Rogers RT5880 | FR4 | Rogers RT5880 | Rogers RO4003C |
| Beam Steering | Yes | No | Yes | |
| Isolation | >25 dB | ~20 dB | 18 dB | 22 dB |
| Gain | 9.2 (5.9GHz); 14.8 dBi (28 GHz) | 6.5 dBi | 7 dBi | 8.5 dBi |
| Efficiency | >85% | >78% | >80% | >82% |
| Application | V2X | DSRC V2V | DSRC | 5G |

## 3. APPLICATION IN AUTONOMOUS VEHICLES

The compact form factor and dual-band support make this antenna suitable for bumper installations and it was installed in an automotive lab. The beamforming ability improves link reliability in NLOS conditions while the DSRC layer ensures short-range safety messaging.

Ray-tracing simulations show improved link reliability compared to horn-based solutions, especially during vehicular turns and lane changes.

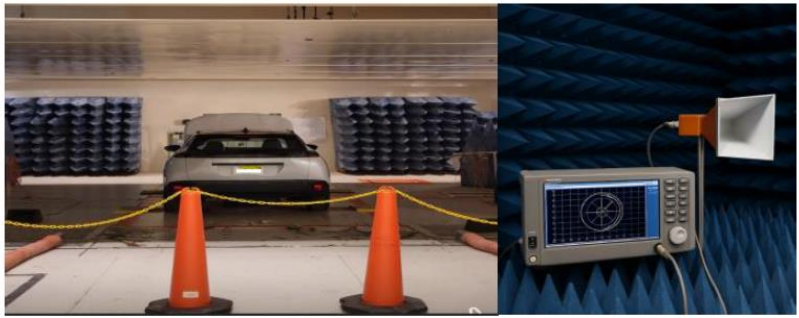

Figure-9: Indoor Antenna Measurement test facility

## 4. DISCUSSION AND FUTURE WORK

The close correlation between simulated and measured results affirms the accuracy of the electromagnetic model and the reliability of the proposed structure for dual-band MIMO operations. The robustness of the antenna during physical testing demonstrates its feasibility for real-time automotive V2X applications. Future work may explore integration with vehicular platforms and over-the-air (OTA) testing under dynamic conditions.

## 5. CONCLUSIONS

This work presented the design, simulation, fabrication, and testing of a high-gain MIMO beamforming patch antenna system that integrates DSRC (5.9 GHz) and mmWave 5G (28 GHz) frequency bands for next-generation autonomous vehicle communication. The dual-band structure was designed using Rogers RT5880 substrate, supporting high-frequency operation with minimal losses and enabling efficient integration into compact vehicular platforms.

The antenna array demonstrated excellent S-parameter characteristics, high radiation efficiency, and significant isolation between elements, making it suitable for multi-element MIMO configurations. Beam steering capabilities were successfully realized through controlled phase shifting, achieving up to ±30° directional control without compromising gain or main lobe fidelity.

Fabrication and empirical measurements validated the simulated performance, with the fabricated prototype achieving gains of 7.1 dBi and 10.5 dBi at 5.9 GHz and 28 GHz, respectively. The results underscore the feasibility of combining V2X communication standards into a unified antenna system for enhanced vehicle-to-everything connectivity.

The proposed antenna system offers a robust solution to meet the requirements of intelligent transportation systems, enabling high-data-rate, low-latency, and directionally controlled communication. Future developments may include integration with vehicular radar, dynamic beam tracking, and AI-assisted tuning to further enhance real-time responsiveness in complex urban environments.

## ACKNOWLEDGEMENTS

I would like to express my sincere gratitude to the Department of Electrical and Computer Engineering at Oakland University for providing the laboratory facilities and technical support by providing simulation tools such as Ansys HFSS, Matlab and Python simulator that enabled the design of the antenna. Special thanks are extended to the automotive communication labs and academic reviewers for their valuable insights during the development phase.